\documentclass[12pt]{article}
\usepackage{amsmath,amssymb,latexsym,graphicx}

\addtocounter{footnote}{2}

\begin{document}

\begin{flushright}
	OUHEP-461 \\
	YITP-03-82 \\
	OIQP-04-03 \\
	hep-th/0312302
\end{flushright}

\vspace{0.5cm}

\begin{center}
{\Large\bf Boson Sea versus Dirac Sea}\\
\vspace{0.2cm}
{\large\bf --General formulation of the boson sea through supersymmetry--}\\
\vspace{0.5cm}
{\large Yoshinobu H}ABARA\\
\vspace{0.2cm}
{\it  Department of Physics, Graduate School of Science,}\\
{\it  Osaka University, Toyonaka, Osaka 560-0043, Japan}\\
\vspace{1cm}
{\large Holger B. N}IELSEN\\
\vspace{0.2cm}
{\it  Niels Bohr Institute, University Copenhagen,}\\
{\it  17 Blegdamsvej Copenhagen \o, Denmark}\\
\vspace{0.3cm}
and\\
\vspace{0.3cm}
{\large Masao N}INOMIYA\footnotemark \\
\vspace{0.2cm}
{\it  Yukawa Institute for Theoretical Physics,}\\
{\it  Kyoto University, Kyoto 606-8502, Japan}
\end{center}

\footnotetext{After April 1, 2004, also working at Okayama Institute for Quantum Physics, Kyoyama-cho 1-9, Okayama-city 700-0015, Japan}

\vspace{0.5cm}

\begin{abstract}
We consider the long standing problem in field theories of bosons that the boson vacuum does not consist of a `sea', unlike the fermion vacuum. We show with the help of supersymmetry considerations that the boson vacuum indeed does also consist of a sea in which the negative energy states are all ``filled", analogous to the Dirac sea of the fermion vacuum, and that a hole produced by the annihilation of one negative energy boson is an anti-particle. Here, we must admit that it is only possible if we allow ---as occurs in the usual formalism anyway--- that the ``Hilbert space" for the single particle bosons is not positive definite. This might be formally coped with by introducing the notion of a double harmonic oscillator, which is obtained by extending the condition imposed on the wave function. This double harmonic oscillator includes not only positive energy states but also negative energy states. We utilize this method to construct a general formalism for a boson sea analogous to the Dirac sea, irrespective of the existence of supersymmetry. The physical result is consistent with that of the ordinary second quantization formalism. We finally suggest applications of our method to the string theories.
\end{abstract}

\newpage

\section{Introduction}

\vspace{0.5cm}

We can consider all present day elementary particle theory to be founded in field theories. One fundamental concept in such theories is expressed by the energy-momentum relation of A. Einstein, 

\begin{align*}
	E^2-\vec{p}^{\> 2}=m^2 \> \Longleftrightarrow \> 
	E=\pm \sqrt{\vec{p}^{\> 2}+m^2}.
\end{align*}

\noindent One implication of this relation is that we have not only positive energy states but also negative energy states. It is difficult to obtain a physical understanding of negative energy states and the implications of their existence. In fact, the existence of particle states of negative energy implies a bottomless energy. For instance, a particle can emit another particle, such as a photon etc., and thereby decay into a lower energy state. It is thus seen that the particle in question cannot exist as a stable state. This has been regarded as one of the fundamental problems in field theories and has vexed elementary particle physicists for many years (for a historical account, see Ref.~\cite{weinberg}). However, in 1930 P. A. M. Dirac succeeded in obtaining a physically consistent particle picture. His well-known interpretation is the following: The fermion vacuum is not an empty state. Rather, all the negative energy states are filled, forming what is called the ``Dirac sea", and thus it is impossible for a positive energy particle to fall into the negative energy states, due to Pauli's exclusion principle. Furthermore, if one negative energy particle disappears, and a hole is thereby created in the Dirac sea, this hole is interpreted as a positive energy particle relative to the surrounding Dirac sea. This hole therefore can be considered a positive energy anti-particle. This surprising prediction was confirmed by the discovery of the positron, the anti-particle of the electron. In this manner, the problem of negative energy states for fermions was solved. Indeed, the chiral anomaly, which is a peculiar quantum effect in fermion theories, is understood and re-derived as one of the physical phenomena resulting from the Dirac sea~\cite{nn}.

What is the situation for bosons? It seems that, in the ordinary understanding of present day particle physics, we are able to obtain consistent boson field theories by re-interpreting creation or annihilation of a negative energy particle as the annihilation or creation of a positive energy particle. The negative energy vacuum in such a theory is empty, like the vacuum of a positive energy solution.

However, we believe that the ordinary boson vacuum theory described above consists basically of a re-reading of the creation and annihilation operators, which are essential tools in quantum theory. This re-reading is nothing but a re-naming of the operators mathematically, and hence we conclude that the ordinary boson theory does not provide a physically satisfactory explanation of the boson vacuum structure. Moreover, in view of supersymmetry, which plays an essential role in recent elementary particle theories, such as superstring theories and supersymmetric grand unified theories, we feel that there is a problem with regard to the fundamental difference between the fermion vacuum and the boson vacuum in the standard treatment: In the fermion vacuum, the negative energy states are totally filled, while the boson vacuum is empty.

It is the purpose of the present article to present an idea about how to construct a ``boson sea"\footnotemark analogous to the Dirac sea by investigating negative energy solutions in detail. In the course of this study, supersymmetry plays an important role. However, we generalize the formalism so that our proposed theory holds in general, even in non-supersymmetric boson theories.

\footnotetext{The idea of a boson sea was proposed by two of the present authors (H.B.N. and M.N.) a few years ago~\cite{nn2}. However, in Ref.~\cite{nn2}, there is used a slightly different notation from the present article with respect to enumerating the extrapolated on negative excitation number states. So $|-n\rangle$ is called $|-n+1\rangle$ in the present article. Then we have two states $|0_+\rangle$ and $|0_-\rangle$ in the present paper with the name $0$. Of course $|0_-\rangle =|-1\rangle \Big|_{\text{Ref.~\cite{nn2}}}$ and $|0_+\rangle =|0\rangle \Big|_{\text{Ref.~\cite{nn2}}}$.}

Often nowadays, even the Dirac sea for fermions is not discussed so much, because most physicists simply work with a formalism of quantum field theory in which one already use creation and annihilation operators even for anti-particles. From the starting point of such a formalism, considering the theory before the filling of the Dirac sea means really to consider a very strange state of the quantum field theory in which we have anti-particles with all possible momenta as many as are allowed. Such a state with such a huge amount of anti-particles is of course far away from any practically producible state and thus can only be of academical and pedagogical interest. So thinking of the world without the fermions filled in the Dirac sea is only of academical and pedagogical interest. But we think that it is often so terrible for obtaining an intuitive feeling for field theories to have the Dirac sea filled that it is really of great help to intuition to be able to avoid this filling in some of our thinking. For example, you cannot even create a fermion with a definite position. But, if the going back to the unfilled Dirac sea has a pedagogical and intuitive purpose, then we should also look for a description of a boson theory in which one could e.g. create a boson with a definite position, and that would presumably be of interest even if the theory to be used by the intuition and pedagogically would suffer from some physical drawbacks such as an indefinite Hilbert space, as we shall see, will be needed.

We shall present in this article to let the inspiration for such an intuitive but indefinite Hilbert space approach be presented as coming from supersymmetry, although the basic ideas were already presented in~\cite{nn,nn2} without using SUSY.

\vspace{0.5cm}

In Section 2, in view of supersymmetry, which asserts the equivalence of bosons and fermions in a certain sense, we investigate the boson vacuum that should be supersymmetric to the fermion vacuum, i.e. the Dirac sea. As a concrete example, we consider the theory of an $N=2$ matter multiplet called a hypermultiplet. In this theory, the condition for the boson vacuum is that it vanishes through the creation of a negative energy particle. This may seem counterintuitive, but in fact this condition is equivalent to that for the fermion vacuum. In Section 3, we introduce a double harmonic oscillator in the first quantized theory. In the ordinary theory, there appear only states with positive energy. However, we extend the condition for the wave function and in so doing obtain negative energy states as well as positive energy states. It is argued that these new negative energy states are precisely what is needed to describe bosons with negative energy. We then show that the boson vacuum is also a kind of filled state, analogous to the fermion vacuum, with the negative energy states filled by bosons. In Section 4, we apply the double harmonic oscillator to the second quantized boson system. We then show that true boson vacuum is obtained from a vacuum analogous to the empty Dirac sea which we associate with a ``sector" calld ``the sector with bottom" by an operation (67) to be seen below. We discuss a new particle picture in which the hole produced by annihilating a negative energy particle is an anti-particle that is observable. We confirm that our theory is physically consistent by considering the energies of states. Section 5 is devoted to a conclusion and a brief overview of possible future developments.

\section{Boson vacuum in a supersymmetric theory}

\vspace{0.5cm}

In the present section, we consider an ideal world in which supersymmetry holds exactly. Then, it is natural to believe that in analogy to the true fermion vacuum the true boson vacuum is a state in which all negative energy states are occupied. To investigate the details of the vacuum structure of bosons needed to realize supersymmetry with the Dirac sea, we utilize the $N=2$ matter multiplet called a hypermultiplet~\cite{sohnius,west}. In fact, we construct the Noether current from the supersymmetric action, and by requiring that the entire system be supersymmetric, we derive the properties of the boson vacuum, while the fermion vacuum is taken to be the Dirac sea.

Hereafter, the Greek indices $\mu ,\nu ,\cdots$ are understood to run from 0 to 3, corresponding to the Minkowski space, and the metric is given by $\eta^{\mu \nu}=diag(+1,-1,-1,-1)$.

\subsection{$N=2$ matter multiplet: Hypermultiplet}

\vspace{0.5cm}

Let us summarize the necessary part of $N=2$ supersymmetric field theory in the free case, in order to reveal the difficulty of making supersymmetric description analogous to a not yet filled Dirac sea. The hypermultiplet is the simplest multiplet that is supersymmetric and involves a Dirac fermion. It is written 

\begin{align}
	\phi =(A_1, A_2;\> \psi ;\> F_1, F_2), 
\end{align}

\noindent where $A_i$ and $F_i\> (i=1,2)$ denote complex scalar fields, and the Dirac field is given by $\psi$. The multiplet (1) transforms under a supersymmetric transformation as 

\begin{align}
	& \delta A_i=2\bar{\xi}_i\psi , \nonumber \\
	& \delta \psi =-i\xi_iF_i-i\gamma^{\mu}\partial_{\mu}\xi_iA_i, 
	\nonumber \\
	& \delta F_i=2\bar{\xi_i}\gamma^{\mu}\partial_{\mu}\psi ,
\end{align}

\noindent where $\gamma^{\mu}$ denotes the four-dimensional gamma matrices, with $\{ \gamma^{\mu},\gamma^{\nu}\} =2\eta^{\mu \nu}$. The commutator of the supersymmetry transformations is given by 

\begin{align}
	[\delta^{(1)},\delta^{(2)}]=2i\bar{\xi}_i^{(1)}\gamma^{\mu}
	\xi_i^{(2)}\partial_{\mu}+2i\bar{\xi}_i^{(1)}\xi_i^{(2)}\delta_Z,
\end{align}

\noindent where $\delta^{(1)}$ and $\delta^{(2)}$ are the supersymmetry transformations associated with $\xi^{(1)}$ and $\xi^{(2)}$ respectively, and $\delta_Z$ represents the $N=2$ supersymmetry transformation with central charge $Z$. Its explicit transformation law takes the form 

\begin{align}
	\left\{ \begin{array}{l}
	\delta_ZA_i=F_i, \\
	\delta_Z\psi =\gamma^{\mu}\partial_{\mu}\psi , \\
	\delta_ZF_i=-\partial_{\mu}\partial^{\mu}A_i.
	\end{array} \right.
\end{align}

\noindent This transformation law satisfies the condition 

\begin{align}
	\delta_Z^2=-\partial_{\mu}\partial^{\mu}=P_{\mu}P^{\mu}=m^2,
\end{align}

\noindent where $m$ denotes the multiplet mass.

A supersymmetric scalar density can be constructed from two hypermultiplets 

\begin{align*}
	& \phi =(A_1, A_2;\> \psi ;\> F_1, F_2), \\
	& \rho =(B_1, B_2;\> \chi ;\> G_1, G_2)
\end{align*}

\noindent as the inner product 

\begin{align*}
	(\bar{\rho}\cdot \phi )\equiv iB_i^{\dagger}F_i-iG_i^{\dagger}A_i
	+2\bar{\chi}\psi ,
\end{align*}

\noindent up to total derivatives. From this scalar density, we obtain the Lagrangian density of the hypermultiplet, 

\begin{align}
	\mathcal{L} & =\frac{i}{2}(\bar{\phi}\cdot \delta_{Z}\phi)
	+\frac{m}{2}(\bar{\phi}\cdot \phi) \nonumber \\
	& =\frac{1}{2}\partial_{\mu}A_i^{\dagger}\partial^{\mu}A_i
	+\frac{1}{2}F_i^{\dagger}F_i+i\bar{\psi}\gamma^{\mu}\partial_{\mu}\psi
	+m\left[\frac{i}{2}A_i^{\dagger}F_i
	-\frac{i}{2}F_i^{\dagger}A_i+\bar{\psi}\psi \right].
\end{align}

\noindent The hermitian form of (6) is given by 

\begin{align}
	\mathcal{L} & =\frac{1}{2}\partial_{\mu}A_i^{\dagger}\partial^{\mu}
	A_i+\frac{1}{2}F_i^{\dagger}F_i+\frac{i}{2}\bar{\psi}\gamma^{\mu}
	\partial_{\mu}\psi -\frac{i}{2}\partial_{\mu}\bar{\psi}\gamma^{\mu}
	\psi +m\left[\frac{i}{2}A_i^{\dagger}
	F_i-\frac{i}{2}F_i^{\dagger}A_i+\bar{\psi}\psi \right].
\end{align}

To derive the Noether currents whose charges generate the supersymmetry transformation, we consider a variation under the supersymmetry transformation (2), 

\begin{align}
	\delta \mathcal{L}=\bar{\xi}_i\partial_{\mu}K_i^{\mu}
	+\partial_{\mu}\bar{K}_i^{\mu}\xi_i,
\end{align}

\noindent where $K_i^{\mu}$ is given by 

\begin{align}
	K_i^{\mu}\equiv \frac{1}{2}(\gamma^{\mu}\gamma^{\nu}\psi 
	\partial_{\nu}A_i^{\dagger}+im\gamma^{\mu}\psi A_i^{\dagger}).
\end{align}

\noindent Thus, the Noether current $J_i^{\mu}$ is written 

\begin{align}
	& \bar{\xi}_iJ_i^{\mu}+\bar{J}_i^{\mu}\xi_i=
	\frac{\delta L}{\delta (\partial_{\mu}\phi)}\delta \phi 
	-\left(\bar{\xi}_iK_i^{\mu}+\bar{K}_i^{\mu}\xi_i\right) \nonumber \\
	& \qquad \qquad \quad =\bar{\xi}_i
	\big(\gamma^{\nu}\gamma^{\mu}\psi \partial_{\nu}
	A_i^{\dagger}-im\gamma^{\mu}\psi A_i^{\dagger}\big)+\big(\partial_{\nu}
	A_i\bar{\psi}\gamma^{\mu}\gamma^{\nu}+imA_i\bar{\psi}\gamma^{\mu}\big)
	\xi_i, \\
	& \quad \Longrightarrow J_i^{\mu}=\gamma^{\nu}\gamma^{\mu}\psi 
	\partial_{\nu}A_i^{\dagger}-im\gamma^{\mu}\psi A_i^{\dagger}.
\end{align}

We could attempt to think of treating the bosons analogous to the fermions by imagining that the creation operators $a^{\dagger}(\vec{k})$ of the anti-bosons were really annihilation operators in some other formulation, but, as we shall see, such an attempt leads to some difficulties. If we could indeed do so, analogously to the pre-Dirac sea filling fermion field consisting solely of annihilation operators, we would write also a boson field in terms of only annihilation operators formally as follows 

\begin{align}
	& A_i(x)=\int \frac{d^3\vec{k}}{\sqrt{(2\pi )^32k_0}}\left\{
	a_{i+}(\vec{k})e^{-ikx}+a_{i-}(\vec{k})e^{ikx}\right\}, \\
	& \psi(x)=\int \frac{d^3\vec{k}}{\sqrt{(2\pi )^32k_0}}\sum_{s=\pm}
	\left\{b(\vec{k},s)u(\vec{k},s)e^{-ikx}+d(\vec{k},s)v(\vec{k},s)e^{ikx}
	\right\}.
\end{align}

\noindent Here, $k_0\equiv \sqrt{\vec{k}^2+m^2}$ is the energy of the particle, and $s\equiv \frac{\vec{\sigma}\cdot \vec{k}}{|\vec{k}|}$ denotes the helicity. Particles with positive and negative energy are described by $a_{i+}(\vec{k}), b(\vec{k},s)$ and $a_{i-}(\vec{k}), d(\vec{k},s)$, respectively. The commutation relations between these field modes are derived as 

\begin{align}
	& \left[a_{i+}(\vec{k}),a_{j+}^{\dagger}(\vec{k}^{\prime})\right]=
	+\delta_{ij}\delta^3(\vec{k}-\vec{k}^{\prime}), \\
	& \left[a_{i-}(\vec{k}),a_{j-}^{\dagger}(\vec{k}^{\prime})\right]=
	-\delta_{ij}\delta^3(\vec{k}-\vec{k}^{\prime}), \\
	& \left\{b(\vec{k},s),b^{\dagger}(\vec{k}^{\prime},s^{\prime})\right\}
	=+\delta_{ss^{\prime}}\delta^3(\vec{k}-\vec{k}^{\prime}), \\
	& \left\{d(\vec{k},s),d^{\dagger}(\vec{k}^{\prime},s^{\prime})\right\}
	=+\delta_{ss^{\prime}}\delta^3(\vec{k}-\vec{k}^{\prime}), 
\end{align}

\noindent with all other pairs commuting or anti-commuting. Note that the right-hand side of the commutation relation (15), for negative energy bosons, has the opposite sign of (14), for positive energy bosons. In the standard context, these creation and annihilation operators have the following interpretations: 

\begin{align*}
	& \left\{ \begin{array}{l}
	a_{i+}(\vec{k})\text{ annihilates a positive energy boson,} \\
	a_{i+}^{\dagger}(\vec{k})\text{ creates a positive energy boson,} \\
	a_{i-}(\vec{k})\text{ annihilates a negative energy boson,} \\
	a_{i-}^{\dagger}(\vec{k})\text{ creates a negative energy boson,} \\
	\end{array} \right. \\
	& \left\{ \begin{array}{l}
	b(\vec{k},s)\text{ annihilates a positive energy fermion,} \\
	b^{\dagger}(\vec{k},s)\text{ creates a positive energy fermion,} \\
	d(\vec{k},s)\text{ annihilates a negative energy fermion,} \\
	d^{\dagger}(\vec{k},s)\text{ creates a negative energy fermion.}
	\end{array} \right.
\end{align*}

\noindent In the ordinary method, recalling that the Dirac sea is the true fermion vacuum, we can use $d^{\dagger}$ as the creation operator and $d$ as the annihilation operator for negative energy fermions. Then, the operators $d^{\dagger}$ and $d$ are re-interpreted as the annihilation operator and creation operator for positive energy holes. In this manner, we obtain the particle picture in the real world. In this procedure, negative energy fermions are regarded as actually existing entities. 

For bosons, in contrast to the fermions, we rewrite (15) as 

\begin{align}
	& \> a_{i-}\equiv \tilde{a}_{i-}^{\dagger}, \quad a_{i-}^{\dagger}
	\equiv \tilde{a}_{i-}, \\
	& \left[\tilde{a}_{i-}(\vec{k}),\tilde{a}_{j-}^{\dagger}
	(\vec{k}^{\prime})\right]=+\delta_{ij}\delta^3
	(\vec{k}-\vec{k}^{\prime}).
\end{align}

\noindent This implies that we can treat negative energy bosons in the same manner as positive energy bosons. Consequently, the true vacua for positive and negative energy bosons, which are denoted $||0_+\rangle$ and $||0_-\rangle$, respectively\footnotemark, are given by 

\footnotetext{In the following, we denote the vacua by, for example in the boson case, $|0_{\pm}\rangle$ in the system of single particle, and $||0_{\pm}\rangle$ in the system with many particles.}

\begin{align*}
	& a_{i+}||0_+\rangle =0, \\
	& \tilde{a}_{i-}||0_-\rangle =0. \tag{$18^{\prime}$}
\end{align*}

\noindent Thus, in the true vacuum, meaning the one on which our experimental world is built, both the negative and positive energy vacua are empty when using the particle $a_{i+}$ and anti-particle $\tilde{a}_{i-}$ annihilation operators respectively. However, in order to have a supersymmetry relation to the analogous negative energy states for the fermions, we would have liked to consider, instead of $||0_-\rangle$, a vacuum so that it were empty with respect to the negative energy bosons described by $a_{i-}$ and $a_{i-}^{\dagger}$. That is to say we would have liked a vacuum obeying $a_{i-}||0_{\text{wanted}}\rangle =0$. Because of ($18^{\prime}$) it is, however, immediately seen that this $||0_{\text{wanted}}\rangle$ cannot exist.

In true nature we should rather be in a situation or a ``sector" in which we have a state with $a_{i-}^{\dagger}||0_-\rangle =\tilde{a}_{i-}||0_-\rangle =0$. It could be called a ``sector with a top" $||0_-\rangle$.

Now, however, we take the point of view that since the game of wanting to study the empty Dirac sea was just for allowing ourselves the intuitive and pedagogical benefits of such a study, we do not have to require that we can achieve this description by a state already presented in the true (and usual) quantum field theory. Rather we shall therefore consider a new sector meaning an extension to the Hilbert space (i.e. the Fock space) corresponding to allowing the number of bosons in the vacuum orbits (single particle states) to be also negative. Actually we have seen in~\cite{nn,nn2} ---and shall review below--- that a formal extension to also allowing negative number of bosons means adding formally a part to the Hilbert space ---which is actually then no more positive definite. The negative number boson state is separated from the unextended part in the sense that one cannot go between the sectors of non-negative and negative numbers of particles with polynomials in creation and annihilation operators.

Perhaps the nicest way of describing this extension is by means of the double harmonic oscillator to be presented in Section 3 below, but let us stress that all we need is a formal extrapolation to also include the possibility of negative numbers of bosons.

\subsection{Supersymmetry invariant vacuum}

\vspace{0.5cm}

As described in Subsection 2.1, when considered in terms of supersymmetry, there is a difference between the boson and fermion pictures. In the present subsection, we give preliminary considerations to the problem determining the nature of a boson sea that would correspond to the Dirac sea for the fermion case. To this end, we impose the natural condition within the supersymmetric theory that the vacuum be supersymmetry invariant. 

We first rewrite the supersymmetry charges $Q_i$ derived from the supersymmetry currents described by Eq.(11) in terms of the creation and annihilation operators as 

\begin{align}
	Q_i & =\int d^3\vec{x} J_i^0(x) \nonumber \\
	& =i\int d^3\vec{k} \sum_{s=\pm} 
	\left\{b(\vec{k},s)u(\vec{k},s)
	a_{i+}^{\dagger}(\vec{k})-d(\vec{k},s)v(\vec{k},s)a_{i-}^{\dagger}
	(\vec{k})\right\}, \\
	\bar{Q}_i & =\int d^3\vec{x} \bar{J}_i^0(x) \nonumber \\
	& =-i\int d^3\vec{k} \sum_{s=\pm} 
	\left\{b^{\dagger}(\vec{k},s)
	\bar{u}(\vec{k},s)a_{i+}(\vec{k})-d^{\dagger}(\vec{k},s)\bar{v}
	(\vec{k},s)a_{i-}(\vec{k})\right\}.
\end{align}

\noindent By applying these charges, the condition for the vacuum to be supersymmetric can be written 

\begin{align}
	Q_i||0\rangle=\bar{Q}_i||0\rangle=0.
\end{align}

\noindent We then decompose the total vacuum into the boson and fermion vacua, $||0_{\pm}\rangle$ and $||\tilde{0}_{\pm}\rangle$, writing 

\begin{align}
	||0\rangle \equiv ||0_+\rangle \otimes ||0_-\rangle \otimes 
	||\tilde{0}_+\rangle \otimes ||\tilde{0}_-\rangle ,
\end{align}

\noindent where $\otimes$ denotes the direct product, and $||\tilde{0}_-\rangle$ is the Dirac sea, given by 

\begin{align*}
	||\tilde{0}_-\rangle =\bigg\{\prod_{\vec{p},s}d^{\dagger}
	(\vec{p},s)\bigg\}||\tilde{0}\rangle .
\end{align*}

\noindent The fermion vacuum described by the above two equations consists of $||\tilde{0}_+\rangle$ and $||\tilde{0}_-\rangle$. Here, $||\tilde{0}_+\rangle$ represents an empty vacuum, annihilated by the ordinary $b$ operator, while $||\tilde{0}_-\rangle$, given by Eq.(23), represents the Dirac sea, which is obtained through application of all $d^{\dagger}$. The condition for the bosonic vacuum reads 

\begin{align}
	& a_{i+}(\vec{k})||0_+\rangle =0, \\
	& a_{i-}^{\dagger}(\vec{k})||0_-\rangle =0.
\end{align}

\noindent It is evident that the vacuum of the positive energy boson $||0_+\rangle$ is the empty one, vanishing under the annihilation operator $a_{i+}$. On the other hand, the vacuum of the negative energy boson $||0_-\rangle$ is defined such that it vanishes under the operator $a_{i+}$ that creates the negative energy quantum. This may seem very strange. 

One could call the strange ``algebra" looked for a ``sector with top", contrary to the more usual creation and annihilation systems which could rather be called ``sectors with a bottom".

In the next section, using the fact that the algebras (14) and (15) constitute that is essentially a harmonic oscillator system with infinitely many degrees of freedom, we investigate in detail the vacuum structure by considering the simplest one-dimensional harmonic oscillator system. In fact, we will find the explicit form of the vacuum $||0_-\rangle$ that is given by a coherent state of the excited states of all the negative energy bosons.

\section{Double harmonic oscillator}

\vspace{0.5cm}

When looking for solutions to the Klein-Gordon equation for energy (and momentum) it is well-known that, we must consider not only the positive energy particles but also the negative energy ones. In the previous section, we found that in order to implement the analogy to the Dirac sea for fermions suggested by supersymmetry, we would have liked to have at our disposal the possibility to organize an analogon of the filling of the Dirac sea (for fermions). This should be organized so as to go from a sector with the naive algebra for $a_{i-}$ meaning as a ``sector with bottom" $a_{i-}||0_{\text{wanted}}\rangle =0$ to an analogon of the filled Dirac sea being a ``sector with top" $a_{i-}^{\dagger}||0_-\rangle =0$. However, so far we are familiar in the description of boson state filling by means of a harmonic oscillator with a ``sector with bottom". In the present section we introduce the concept of a ``sector with top" as an extension of the harmonic oscillator spectrum to a negative energy sector. Thereby we have to extend the ordinary meaning of the wave function (in this case for the harmonic oscillator). Performing this we find that the vacuum of the negative energy sector leads to a ``boson sea", corresponding to the Dirac sea of fermions.

\subsection{Analytic wave function of the harmonic oscillator}

\vspace{0.5cm}

As is well known, the eigenfunction $\phi (x)$ of a one-dimensional Schr\"{o}dinger equation in the usual treatment should satisfy the square integrability condition, 

\begin{align}
	\int_{-\infty}^{+\infty}dx \> |\phi (x)|^2<+\infty .
\end{align}

\noindent If we apply this condition to a one-dimensional harmonic oscillator, we obtain as the vacuum solution only the empty one satisfying (24), 

\begin{align*}
	a_+|0\rangle =0.
\end{align*}

\noindent Thus, we are forced to extend the condition for physically allowed wave functions in order to obtain ``boson sea" analogous to the Dirac sea. In fact we extend the condition (26), replacing it by the condition under which, when we analytically continuate $x$ to the entire complex plane, the wave function $\phi (x)$ is analytic and only an essential singularity is allowed as $|x|\! \to \! \infty$.

The one-dimensional harmonic oscillator is given by 

\begin{align}
	\left( -\frac{1}{2}\frac{d^2}{dx^2}+\frac{1}{2}x^2 \right) 
	\phi (x)=E\phi (x).
\end{align}

\noindent Under the condition that the wave function $\phi (x)$ be analytic, the function $H_n(x)e^{-\frac{1}{2}x^2}$ is obviously a solution, where $H_n(x)$ denotes the Hermite polynomial. In order to find further solutions we assume the form 
\begin{align}
	\phi (x)=f(x)e^{\pm \frac{1}{2}x^2}, 
\end{align}

\noindent where $f(x)$ is determined below. Inserting (28) into (27), we obtain the following equation for $f(x)$: 

\begin{align}
	\frac{f^{\prime \prime}(x)}{f(x)}\pm \frac{2xf^{\prime}(x)}{f(x)}=
	-2\left(E\pm \frac{1}{2}\right).
\end{align}

\noindent We then assume that for large $|x|$ the first term on the left-hand side dominates the second term. With this, we obtain from Eq.(29) the simpler equation 

\begin{align}
	\frac{d\log f(x)}{d\log x}=\mp E-\frac{1}{2}.
\end{align}

\noindent The right-hand side of the Eq.(30) is constant, and we denote it by $n\in \mathcal{R}$: 

\begin{align}
	n\equiv \mp E-\frac{1}{2}\in \mathcal{R}.
\end{align}

\noindent We thus find that the large $|x|$ behavior of $f(x)$ is given by 

\begin{align}
	f(x)\sim x^n.
\end{align}

\noindent If $n$ is negative or a non-integer then $f(x)$ may have a cut when $x$ is analytically continuated to the whole complex plane. This contradicts our assumption. Therefore, it should be the case that 

\begin{align*}
	n\in \mathcal{Z}_+\cup \{0\}; 
\end{align*}

\noindent i.e. $n$ is a positive integer or zero ($n=0,1,2,3,\cdots$). Thus, the energy spectrum is given by 

\begin{align}
	E=\mp n\pm \frac{1}{2}=\mp \left( n+\frac{1}{2}\right) , \qquad 
	n\in \mathcal{Z}_{+}\cup \{0\}.
\end{align}

\noindent To further simplify the situation, in the case of the upper sign here, we make the replacement $n\! \to \! -n-1$. The energy spectrum can then be written in the form 

\begin{align}
	E=n+\frac{1}{2}, \qquad n\in \mathcal{Z}.
\end{align}

\noindent That is to say, in addition to the usual solutions for which $n=0,1,2,\cdots$, the negative series $n=-1,-2,\cdots$ is also allowed.

In view of the above arguments regarding the energy spectrum, the eigenfunction for a negative integer $n$ can be written 

\begin{align*}
	\phi (x)=f(x)e^{+\frac{1}{2}x^2}.
\end{align*}

\noindent We determine the explicit form of $f(x)$ in the following: As is well known, the ordinary harmonic oscillator is characterized by the energy eigenvalues (34) with $n\geq 0$ and the eigenfunctions 

\begin{align}
	& \phi_n(x)=\left( \sqrt{\pi}2^nn!\right)^{-\frac{1}{2}}H_n(x)
	e^{-\frac{1}{2}x^2}, \nonumber \\
	& E=n+\frac{1}{2},\qquad n=0,1,2,\cdots .
\end{align}

\noindent Now, we see that the negative energy sector is described by Eq.(34) with $n\leq -1$ or, more precisely, 

\begin{align}
	E=-\left( n+\frac{1}{2}\right) ,\qquad n=0,1,2,\cdots .
\end{align}

\noindent The negative energy sector eigenfunctions, although satisfying the same equation [i.e., Eq.(24)] as the positive energy eigenfunctions, are found to be 

\begin{align*}
	\phi (x)=f(x)e^{+\frac{1}{2}x^2}.
\end{align*}

\noindent This implies that the harmonic oscillator in the negative energy sector is described by simply replacing $x$ by $ix$ in all the results for the positive energy sector. Thus, the eigenfunctions in the negative energy sector can be obtained through the replacement 

\begin{align}
	& \left( -\frac{1}{2}\frac{d^2}{d(ix)^2}+\frac{1}{2}(ix)^2 \right) 
	\phi (x)=E\phi (x) \nonumber \\
	& \quad \Longleftrightarrow \left( -\frac{1}{2}\frac{d^2}{dx^2}
	+\frac{1}{2}x^2 \right) \phi (x)=-E\phi (x).
\end{align}

\noindent The solutions read 

\begin{align}
	& -E=-\left( n+\frac{1}{2}\right) ,\qquad n=0,1,2,\cdots \\
	& \phi_{-n}(x)=A_nH_n(ix)e^{-\frac{1}{2}(ix)^2}=A_nH_n(ix)
	e^{+\frac{1}{2}x^2},
\end{align}

\noindent where $A_n$ denotes a normalization factor to be determined later.

It should be noted that the Hermite polynomial $H_n(x)$ is either an even or odd function of $x$, so that $H_n(ix)$ in Eq.(39) is either purely real or purely imaginary. In the case that it is purely imaginary, the overall factor of $i$ can be absorbed into the normalization factor $A_n$. Thus, the eigenfunctions of the negative energy sector $\phi _{-n}(x)$ can be treated as a real function. Corresponding to the replacement $x\! \to \! ix$, the creation and annihilation operators are transformed as 

\begin{align}
	a_+=\frac{1}{\sqrt{2}}\left( x+\frac{d}{dx}\right) & \to 
	\frac{i}{\sqrt{2}}\left( x-\frac{d}{dx}\right)\equiv a_-^{\dagger}, 
	\nonumber \\
	a_+^{\dagger}=\frac{1}{\sqrt{2}}\left( x-\frac{d}{dx}\right) & \to 
	\frac{i}{\sqrt{2}}\left( x+\frac{d}{dx}\right)\equiv a_-, \\
	& \nonumber \\
	[a_+,a_+^{\dagger}]=+1 & \to [a_-,a_-^{\dagger}]=-1.
\end{align}

\noindent Here, $a_+$ and $a_+^{\dagger}$ are ordinary operators in the positive energy sector which annihilate and create positive energy quanta respectively, and $a_-$ and $a_-^{\dagger}$ are operators in the negative energy sector which, as we will see below, create and annihilate negative energy quanta respectively. It should be noted that the replacement $x\! \to \! ix$ causes the hermite operators $\hat{x}=x$ and $\hat{p}=-i\frac{d}{dx}$ to be transformed to anti-hermite operators. That is to say, the representation is different between the positive and negative energy sector in the operator formalism. At this stage, we can explicitly check the unusual property mentioned above of the negative energy sector, as explained below. The exotic properties that the vacuum is, in our usual words, annihilated by the creation operator $a_-^{\dagger}$ and the annihilation operator $a_-$ excites states are confirmed as 

\begin{align*}
	\text{(vacuum)} \quad & -i\sqrt{2}a_-^{\dagger}\phi _{-0}(x)
	=\left( x-\frac{d}{dx}\right)e^{+\frac{1}{2}x^2}=0, \\
	\text{(excitations)} \quad 
	& \phi_{-1}(x)=2xe^{+\frac{1}{2}x^2}=-i\sqrt{2}a_-\phi _{-0}(x)
	=\left( x+\frac{d}{dx}\right)e^{+\frac{1}{2}x^2}, \\
	& \phi_{-2}(x)=(4x^2+2)e^{+\frac{1}{2}x^2}=-i\sqrt{2}
	a_-\phi _{-1}(x)=\left( x+\frac{d}{dx}\right)^2e^{+\frac{1}{2}x^2}, \\
	& \quad \vdots
\end{align*}

\noindent Also we can confirm that 

\begin{align*}
	\phi_{-n}(x)=a_-^{\dagger}\phi_{-n+1}(x), \quad \text{for}\quad 
	n\geq 1. 
\end{align*}

\noindent Namely, we can consider that $a_-$ creates a negative energy quantum and $a_-^{\dagger}$ annihilates a negative energy quantum. In the remainder of the paper, we rewrite 

\begin{align*}
	& a_- \to ia_-, \\
	& a_-^{\dagger} \to ia_-^{\dagger}, 
\end{align*}

\noindent since the imaginary unit $i$ is not essential in our argument.

It is useful to summarize the various results obtained to this point in operator form. We write each vacuum and excited state in the positive and negative energy sectors, respectively, as 

\begin{align}
	& \phi_{+0}(x)=e^{-\frac{1}{2}x^2}\simeq |0_+\rangle , \\
	& \phi_{-0}(x)=e^{+\frac{1}{2}x^2}\simeq |0_-\rangle , \\
	& \phi_n(x)\simeq |n\rangle , \qquad n\in \mathcal{Z}-\{0\}.
\end{align}

\noindent The actions of each operator $a_{\pm}$ and $a_{\pm}^{\dagger}$ acting on various states are summarized in Table 1. Here, overall factors are omitted for simplicity.

\begin{center}
	\begin{tabular}{|l|l|l|} \hline
	& \multicolumn{2}{|c|}{$n=0,1,2,\cdots$} \\ \hline
	& positive energy sector & negative energy sector \\
	& =``sector with bottom" & =``sector with top" \\ \hline
	spectrum & $E=+(n+\frac{1}{2})$ & $E=-(n+\frac{1}{2})$ \\
	& $a_+^{\dagger}|n\rangle =|n+1\rangle$ & $a_-|-n\rangle 
	=|-n-1\rangle$ \\
	& $a_+|n+1\rangle =|n\rangle$ & $a_-^{\dagger}|-n-1\rangle 
	=|-n\rangle$ \\
	vacuum & $a_+|0_+\rangle =0$ & $a_-^{\dagger}|0_-\rangle =0$ \\ \hline
	\multicolumn{3}{c}{} \\
	\multicolumn{3}{c}{Table 1: The actions of $a_{\pm}=(x+\frac{d}{dx})$ 
	and $a_{\pm}^{\dagger}=(x-\frac{d}{dx})$} \\
	\multicolumn{3}{c}{on various states.$\qquad \qquad \qquad \qquad 
	\quad$} \\
	\end{tabular}
\end{center}

The important point here is that there exists a gap between the positive and negative sectors. Suppose that we write the states in order of their energies as 

\begin{figure}[h]
	\begin{center}
	\begin{picture}(360,50)
	\put(40,20){$\cdots$}
	\put(65,23){$\xrightarrow[]{a^{\dagger}}$}
	\put(65,14){$\xleftarrow[\> a \>]{}$}
	\put(85,18){$|\! -\! 1\rangle$}
	\put(115,23){$\xrightarrow[]{a^{\dagger}}$}
	\put(115,14){$\xleftarrow[\> a \>]{}$}
	\put(135,18){$|0_-\rangle$}
	\put(160,23){$\xrightarrow[]{a^{\dagger}}$}
	\put(180,18){$0$}
	\put(190,14){$\xleftarrow[\> a \>]{}$}
	\put(210,18){$|0_+\rangle$}
	\put(235,23){$\xrightarrow[]{a^{\dagger}}$}
	\put(235,14){$\xleftarrow[\> a \>]{}$}
	\put(255,18){$|\! +\! 1\rangle$}
	\put(285,23){$\xrightarrow[]{a^{\dagger}}$}
	\put(285,14){$\xleftarrow[\> a \>]{}$}
	\put(310,20){$\cdots$.}
	\end{picture}
	\end{center}
	\label{sequence}
\end{figure}%

\noindent As usual, the operators causing transitions in the right and left directions are $a_{\pm}^{\dagger}$ and $a_{\pm}$, respectively. However, between the two vacua $|0_-\rangle$ and $|0_+\rangle$ there is a ``wall" of the classical number $0$, and due to its presence, these two vacua cannot be transformed into each other under the operations of $a_{\pm}$ and $a_{\pm}^{\dagger}$, because the states $|0_{\pm}\rangle$ cannot be obtained by means of acting the operators $a_+^{\dagger}$ and $a_-$ on the classical number $0$. In going to the second quantized theory with interactions, there appears to be the possibility of such a transition. However, as discussed in detail in Subsection 3.3 and Section 4, it turns out that the usual polynomial interactions do not induce such a transition.

To end this subsection, we define the inner product of states. As explained above, there exists a gap such that no transition between the two sectors can take place. Thus, we can define the inner product of only states in the same sector. The inner product that in the positive energy sector provides the normalization condition is, as usual, given by 

\begin{align}
	\langle n|m\rangle & \equiv \int_{-\infty}^{+\infty}dx\> 
	\phi_n^{\dagger}(x)\phi_m(x)=\delta_{nm}, \qquad n,m=0_+,1,2,\cdots .
\end{align}

\noindent However, the eigenfunctions in the negative energy sector are obtained as Eq.(39), and consequently a definition of the inner product similar to Eq.(45) leads to divergence in this case if we choose the integration region as $-\infty <x<+\infty$. 

Because of the above-stated problem, we propose  a path of integration such that, after analytic continuation of the wave functions to the whole complex $x$ plane, the integration is convergent, since we impose the condition that the wave functions are analytic. Then recalling that the wave function is real even in the negative energy sector, we define the inner product in terms of a path $\Gamma$, which we make explicit subsequently: 

\begin{align}
	\langle n|m\rangle \equiv -i\int_{\Gamma}dx\> \phi_n^{\dagger}(x)
	\phi_m(x), \qquad n,m=0_-,-1,-2,\cdots
\end{align}

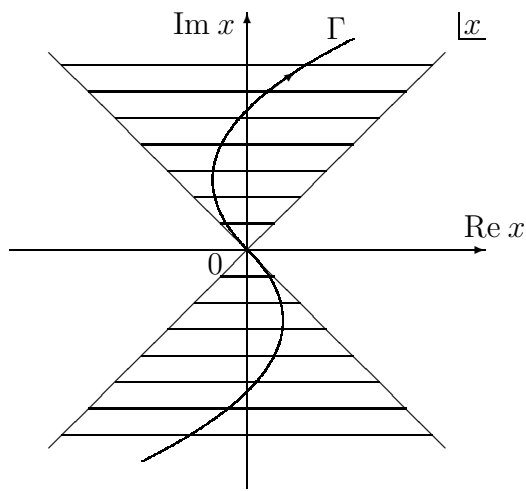
\begin{figure}[ht]
	\begin{center}
	\begin{picture}(180,180)
	\put(0,90){\vector(1,0){180}}
	\put(90,0){\vector(0,1){180}}
	\put(170,170){\line(1,0){10}}
	\put(170,170){\line(0,1){10}}
	\put(172,172){$x$}
	\put(75,81){$0$}
	\put(172,95){Re$\> x$}
	\put(62,172){Im$\> x$}
	\put(15,15){\line(1,1){150}}
	\put(15,165){\line(1,-1){150}}
	\put(80,100){\line(1,0){20}}
	\put(70,110){\line(1,0){40}}
	\put(60,120){\line(1,0){60}}
	\put(50,130){\line(1,0){80}}
	\put(40,140){\line(1,0){100}}
	\put(30,150){\line(1,0){120}}
	\put(20,160){\line(1,0){140}}
	\put(80,80){\line(1,0){20}}
	\put(70,70){\line(1,0){40}}
	\put(60,60){\line(1,0){60}}
	\put(50,50){\line(1,0){80}}
	\put(40,40){\line(1,0){100}}
	\put(30,30){\line(1,0){120}}
	\put(20,20){\line(1,0){140}}
	\qbezier(90,90)(50,130)(130,170)
	\qbezier(90,90)(130,50)(50,10)
	\put(120,170){$\Gamma$}
	\put(98,150){\vector(4,3){10}}
	\end{picture}
	\end{center}
	\caption[path $\Gamma$]{Path $\Gamma$ for which the inner product (46) 
	converges.}
	\label{gamma}
\end{figure}%

\noindent Here, it is understood that the complex conjugation yielding $\phi^{\dagger}$ is taken so that the inner product is invariant under deformations of the path $\Gamma$ within the same topological class (as for definition, see below). The path for which the integration (46) is convergent can be chosen as any path going through the two lined regions $\frac{\pi}{4}<\arg x<\frac{3\pi}{4}$ and $\frac{5\pi}{4}<\arg x<\frac{7\pi}{4}$, as shown in Fig.~\ref{gamma}. In fact, also in the positive energy sector, the inner product (45) is path-independent within the same topological class going through the two unlined regions $-\frac{\pi}{4}<\arg x<\frac{\pi}{4}$ and $\frac{3\pi}{4}<\arg x<\frac{5\pi}{4}$. For simplicity, we choose hereafter the simplest path, $\Gamma =(-i\infty,+i\infty )$, i.e. the imaginary axis. Then, the normalization condition reads 

\begin{align}
	\langle n|m\rangle & =-i\int_{\Gamma}dx\> \phi_n^{\dagger}(x)
	\phi_m(x), \qquad n,m=0_-,-1,-2,\cdots \nonumber \\
	& \qquad \qquad \quad \text{where} \quad \phi_n^{\dagger}(x)\equiv 
	\left[ \phi_n(x^{\ast})\right]^{\ast}, \nonumber \\
	\langle n|m\rangle & =-i(-1)^{n^{\prime}}\int_{-i\infty}^{+i\infty}dx
	\> \phi_{n^{\prime}}(ix)\phi_{m^{\prime}}(ix), \qquad 
	n^{\prime},m^{\prime}=0_+,1,2,\cdots 
	\nonumber \\
	& =(-1)^{n^{\prime}}\int_{-\infty}^{+\infty}dy\> \phi_{n^{\prime}}(y)
	\phi_{m^{\prime}}(y) \nonumber \\
	& =(-1)^{n^{\prime}}\delta_{n^{\prime}m^{\prime}} \nonumber \\
	& =(-1)^{n}\delta_{nm}.
\end{align}

\noindent Equation (47) determines the normalization factor $A_n$ of the eigenfunction (39): 

\begin{align}
	\phi_{n}(x)=\left( \sqrt{\pi}2^{|n|}|n|!\right)^{-\frac{1}{2}}i^{|n|}
	H_{|n|}(ix)e^{+\frac{1}{2}x^2}.
\end{align}

Some comments are in order. When we performed the integration after the analytic continuation, we introduced the variable $y\equiv ix$. For the one-dimensional harmonic oscillator, $x$ denotes the spatial coordinate. Therefore $y$ is purely imaginary, and hence it does not possess the physical meaning of a spatial coordinate. However, when we apply the analytic wave function to an infinite-dimensional harmonic oscillator as the second quantized theory, the integration variable is the wave function $\phi$, and thus the change of variable to $\omega$, defined as $\omega \equiv i\phi$, does not affect the physical content. Therefore, the definition of the inner product in the negative energy sector is not essentially different from that of the positive energy sector, except for the result of the alternating signature $(-1)^{n}\delta_{nm}$, which is, however, very crucial.

\subsection{Fermionic harmonic oscillator}

\vspace{0.5cm}

We have derived the explicit form of the eigenfunction in the negative energy sector. However, it should be clarified that the boson vacuum $|0_-\rangle \simeq e^{+\frac{1}{2}x^2}$ in the negative energy sector is really a ``boson sea", corresponding to the Dirac sea in the case of fermions. Here, we recall the concept of a ``sea" by considering the fermionic harmonic oscillator as the first quantized theory.

The one-dimensional fermionic harmonic oscillator is defined in terms of the Grassmann odd operators that satisfy the anti-commutator 

\begin{align}
	\{b,b^{\dagger}\}=1.
\end{align}

\noindent The representations for these operators are expressed in terms of the real Grassmann variable $\theta$ as 

\begin{align}
	b=\theta , \quad b^{\dagger}=\frac{d}{d\theta}.
\end{align}

\noindent This follows from the fact that $b$ and $b^{\dagger}$ are Grassmann odd and they act from the left in the functional space consisting of the Grassmannian functions. The Hamiltonian for the fermionic harmonic oscillator reads 

\begin{align}
	H=\frac{1}{2}[b^{\dagger},b]=b^{\dagger}b-\frac{1}{2}=N-\frac{1}{2}, 
\end{align}

\noindent where $N=b^{\dagger}b$ is the particle number operator in the second quantized theory. The solutions of the Schr\"{o}dinger equation 

\begin{align}
	\left(b^{\dagger}b-\frac{1}{2}\right)|\tilde{n}_+\rangle 
	=E_n|\tilde{n}_+\rangle 
	\Longrightarrow \left(\frac{d}{d\theta}\theta-\frac{1}{2}\right)
	\rho_n(\theta)=E_n\rho_n(\theta)
\end{align}

\noindent have the following forms: 

\begin{align}
	& E_n=n-\frac{1}{2},\qquad n=0,1 \\
	& \rho_0(\theta)=\theta \simeq |\tilde{0}_+\rangle , \\
	& \rho_1(\theta)=1 \simeq |\tilde{1}_+\rangle , \\
	& \text{vacuum} \quad b|\tilde{0}_+\rangle =0.
\end{align}

\noindent Here, the normalization condition 

\begin{align}
	\int d\theta \> \rho_n^{\ast}(\theta)\left(b+b^{\dagger}\right)
	\rho_m(\theta)=\delta_{nm}
\end{align}

\noindent has been used. Note that the vacuum $|0_+\rangle$ is annihilated by the annihilation operator $b$.

In the argument presented in Section 2 for the Dirac sea, in the negative energy sector, the excited states have the same energies as in the positive energy sector when we exchange the operators $b$ and $b^{\dagger}$. With this replacement, the Hamiltonian reads 

\begin{align}
	H=\frac{1}{2}[b,b^{\dagger}]=bb^{\dagger}-\frac{1}{2},
\end{align}

\noindent and consequently the Schr\"{o}dinger equation is changed to 

\begin{align}
	\left(bb^{\dagger}-\frac{1}{2}\right)|\tilde{n}_-\rangle 
	=E_n|\tilde{n}_-\rangle 
	\Longrightarrow \left(\theta \frac{d}{d\theta}-\frac{1}{2}\right)
	\chi_n(\theta)=E_n\chi_n(\theta).
\end{align}

\noindent The solutions of Eq.(59) are given by 

\begin{align}
	& E_n=n-\frac{1}{2},\qquad n=0,1 \\
	& \chi_0(\theta)=1 \simeq |\tilde{0}_-\rangle , \\
	& \chi_1(\theta)=\theta \simeq |\tilde{1}_-\rangle , \\
	& \text{vacuum} \quad b^{\dagger}|\tilde{0}_-\rangle =0,
\end{align}

\noindent where we have used the same normalization condition as for the positive energy sector (57). Here, it should be noted that the vacuum $|\tilde{0}_-\rangle$ in the negative energy sector is annihilated by the creation operator $b^{\dagger}$. This implies that the vacuum $|\tilde{0}_-\rangle$ has precisely the properties postulated to exist for the Dirac sea. In order to understand more precisely the reason why the vacuum $|\tilde{0}_-\rangle$ in the negative energy sector is naturally characterized by the concept of a ``sea", we write the states in terms of the Grassmann variable: 

\begin{align}
	& |\tilde{0}_+\rangle \simeq \theta ,\quad |\tilde{0}_-\rangle 
	\simeq 1 \nonumber \\
	& \Longrightarrow |\tilde{0}_-\rangle =b^{\dagger}|\tilde{0}_+\rangle .
\end{align}

\noindent This equation shows that the vacuum $|\tilde{0}_-\rangle$ in the negative energy sector is obtained by exciting one quantum of the empty vacuum $|\tilde{0}_+\rangle$. If we employ this fact in the second quantized theory, i.e. an infinitely many body system, the vacuum in our picture is equivalent to the Dirac sea.

\subsection{Boson vacuum in the negative energy sector}

\vspace{0.5cm}

We are ready to clarify the meaning of which the vacuum in the negative energy sector of bosons forms a sea, applying the foregoing arguments to the case of bosons. 

The vacua $|0_+\rangle$ and $|0_-\rangle$ in the positive and negative energy sectors are 

\begin{align}
	& |0_+\rangle \simeq e^{-\frac{1}{2}x^2}, \\
	& |0_-\rangle \simeq e^{+\frac{1}{2}x^2}.
\end{align}

\noindent In order to demonstrate how $|0_-\rangle$ represents a sea, we derive a relation between the two vacua (65) and (66) analogous to that in the fermion case. In fact, by comparing the explicit functional forms of each vacuum, we easily find the relation 

\begin{align}
	& e^{+\frac{1}{2}x^2}=e^{x^2}\cdot e^{-\frac{1}{2}x^2}, \qquad 
	e^{x^2}=e^{\frac{1}{2}(a+a^{\dagger})^2} \nonumber \\
	& \Longrightarrow 
	|0_-\rangle =e^{\frac{1}{2}(a+a^{\dagger})^2}|0_+\rangle .
\end{align}

\noindent This relation is preferable for bosons for the following reason. In the fermion case, described by (64), due to the exclusion principle, the Dirac sea is obtained by exciting only one quantum of the empty vacuum. Contrastingly, because in the boson case there is no exclusion principle, the vacuum $|0_-\rangle$ in the negative energy sector is constructed as a sea by exciting all even number of quanta, i.e. an infinite number of quanta. 

Although the basic idea of the boson sea is described above, in this form, it is limited to the first quantized theory. We extend our argument to the second quantized theory in the next section. We present a new particle picture employing our result that the true boson vacuum is similar to the true fermion vacuum, where a hole appearing through the annihilation of one negative energy particle is interpreted as an anti-particle.

\section{Boson sea}

\vspace{0.5cm}

In the previous section, we investigated a one-dimensional harmonic oscillator in detail. This harmonic oscillator describes a one-particle energy state, and it can have a negative number of excitations and be in what we call ``the sector with top". We can apply this result to the second quantization for a complex scalar field, which is simply a system of infinite-dimensional harmonic oscillators. The negative energy quantum describes one negative energy particle, and the vacuum in the negative energy sector is described as the state in which an infinite number of negative energy particles fill all states. 

In the present section, we investigate the boson vacuum structure in detail. We find that by appropriately defining the energy-momentum operator, our method yields the result that all physical phenomena are characterized by positive energy, even in the negative energy sector. Thus it provides a picture that is consistent with the ordinary one derived from a re-definition of the operators (18) and (19). It is important that we can extend all the concepts of the one-dimensional harmonic oscillator to the infinite-dimensional case, since, as is well known in the second quantization, the number of independent (i.e. non-interacting) harmonic oscillators becomes infinite. 

\vspace{0.5cm}

To begin, we summarize the results obtained in Section 2. To simplify the points, we omit the well-known fermion vacuum, the Dirac sea, and furthermore we consider only one boson, $A(x)$, because two complex scalar fields are needed to preserve the supersymmetry, which does not play an essential role in the present section. If we wish to apply the content of the present section to a supersymmetric hypermultiplet, we simply attach the suffices $i=1,2$ to all the scalar fields.

The mode expansion and the commutation relations are given by 

\begin{align}
	& A(x)=\int \frac{d^3\vec{k}}{\sqrt{(2\pi )^32k_0}}\left\{
	a_+(\vec{k})e^{-ikx}+a_-(\vec{k})e^{ikx}\right\}, \\
	& \left[a_+(\vec{k}),a_+^{\dagger}(\vec{k}^{\prime})\right]=
	+\delta^3(\vec{k}-\vec{k}^{\prime}), \nonumber \\
	& \left[a_-(\vec{k}),a_-^{\dagger}(\vec{k}^{\prime})\right]=
	-\delta^3(\vec{k}-\vec{k}^{\prime}).
\end{align}

\noindent As is noted, the commutation relations for the positive energy sector, $a_+(\vec{k})$, and the negative energy sector, $a_-(\vec{k})$, have opposite signs on the right-hand side. The interpretation of these operators and their conjugates are as follows: 

\begin{align*}
	& \left\{ \begin{array}{l}
	a_+(\vec{k})\text{ annihilates a positive energy boson,} \\
	a_+^{\dagger}(\vec{k})\text{ creates a positive energy boson,} \\
	\end{array} \right. \\
	& \left\{ \begin{array}{l}
	a_-(\vec{k})\text{ annihilates a negative energy boson,} \\
	a_-^{\dagger}(\vec{k})\text{ creates a negative energy boson.} \\
	\end{array} \right.
\end{align*}

\noindent The vacuum is expressed as a direct product of those for the positive and negative energy sectors: 

\begin{align}
	||0\rangle_{\text{boson}}=||0_+\rangle \otimes ||0_-\rangle .
\end{align}

\noindent The conditions for $||0_+\rangle$ and $||0_-\rangle$ are given as 

\begin{align}
	& a_+(\vec{k})||0_+\rangle =0, \\
	& a_-^{\dagger}(\vec{k})||0_-\rangle =0,
\end{align}

\noindent respectively. It should be noted that the positive energy sector vacuum $||0_+\rangle$ vanishes under the action of the annihilation operator $a_+$, and $||0_+\rangle$ turns out to be empty, while the negative energy sector vacuum $||0_-\rangle$ vanishes under the action of the ``creation" operator $a_-^{\dagger}$. This ends the summary of Section 2.

\vspace{0.5cm}

Firstly, we clarify the properties of the unfamiliar vacuum $||0_-\rangle$ in the negative energy sector, using the result of Section 3. To this end, we study the details of the infinite-dimensional harmonic oscillator, which is identical to a system of a second quantized complex scalar field. The representation of the algebra (69) that is formed by $a_+,a_-$ and their conjugate operators is expressed as 

\begin{align}
	& a_+(\vec{k})=\left( A(\vec{k})+\frac{\delta}{\delta A(\vec{k})}
	\right), \quad a_+^{\dagger}(\vec{k})=\left( A(\vec{k})
	-\frac{\delta}{\delta A(\vec{k})}\right), \\
	& a_-(\vec{k})=i\left( A(\vec{k})+\frac{\delta}{\delta A(\vec{k})}
	\right),\quad a_-^{\dagger}(\vec{k})=i\left( A(\vec{k})
	-\frac{\delta}{\delta A(\vec{k})}\right).
\end{align}

\noindent The Hamiltonian and Schr\"{o}dinger equation of this system as the infinite-dimensional harmonic oscillator read 

\begin{align}
	& H=\int \frac{d^3\vec{k}}{(2\pi )^3}\left\{ -\frac{1}{2}
	\frac{\delta^2}{\delta A^2(\vec{k})}+\frac{1}{2}A^2(\vec{k})\right\} 
	, \\
	& H\Phi [A]=E\Phi [A].
\end{align}

\noindent Here, $\Phi [A]$ denotes a wave functional of the wave function $A(\vec{k})$. We are now able to write an explicit wave functional for the vacua of the positive and negative enegy sectors: 

\begin{align}
	& ||0_+\rangle \simeq \Phi_{0_+}[A]=e^{-\! \frac{1}{2} \int \! 
	\frac{d^3\vec{k}}{(2\pi )^3}A^2(\vec{k})}, \\
	& ||0_-\rangle \simeq \Phi_{0_-}[A]=e^{+\! \frac{1}{2} \int \! 
	\frac{d^3\vec{k}}{(2\pi )^3}A^2(\vec{k})}.
\end{align}

\noindent We can find a relation between these two vacua via Eq.(67): 

\begin{align}
	||0_-\rangle & =e^{\int \! \frac{d^3\vec{k}}{(2\pi )^3}A^2(\vec{k})}
	||0_+\rangle \nonumber \\
	& =e^{-\frac{1}{2}\! \int \! \frac{d^3\vec{k}}{(2\pi )^3}
	\left\{ a_-(\vec{k})+a_-^{\dagger}(\vec{k})\right\}^2}||0_+\rangle .
\end{align}

\noindent From this equation, we see that the negative energy vacuum $||0_-\rangle$ is a coherent state constructed from the empty vacuum $||0_+\rangle$ of the positive energy sector by creating all the even number negative energy bosons through the action of $a_-^{\dagger}(\vec{k})$. In this sense, $||0_-\rangle$ is the sea in which all the negative energy boson states are filled.

\begin{figure}[ht]
	\begin{center}
	\begin{picture}(200,160)
	\put(35,0){\line(0,1){140}}
	\put(50,0){\line(0,1){140}}
	\put(150,0){\line(0,1){140}}
	\put(165,0){\line(0,1){140}}
	\put(93,67){\Large $\otimes$}
	\put(40,-10){\Large $\vdots$}
	\put(40,140){\Large $\vdots$}
	\put(155,-10){\Large $\vdots$}
	\put(155,140){\Large $\vdots$}
	\put(-25,130){\scriptsize positive energy}
	\put(-5,120){\scriptsize sector}
	\put(185,20){\scriptsize negative energy}
	\put(205,10){\scriptsize sector}
	\multiput(35,75)(0,10){7}{\line(1,0){15}}
	\multiput(150,5)(0,10){7}{\line(1,0){15}}
	\multiput(0,70)(10,0){9}{\line(1,0){5}}
	\multiput(115,70)(10,0){9}{\line(1,0){5}}
	\put(53,73){\scriptsize $0_+$}
	\put(53,83){\scriptsize $+1$}
	\put(53,93){\scriptsize $+2$}
	\put(53,103){\scriptsize $+3$}
	\put(53,113){\scriptsize $+4$}
	\put(53,123){\scriptsize $+5$}
	\put(53,133){\scriptsize $+6$}
	\put(168,63){\scriptsize $0_-$}
	\put(168,53){\scriptsize $-1$}
	\put(168,43){\scriptsize $-2$}
	\put(168,33){\scriptsize $-3$}
	\put(168,23){\scriptsize $-4$}
	\put(168,13){\scriptsize $-5$}
	\put(168,3){\scriptsize $-6$}
	\put(205,67){wall of $0$ (zero)}
	\end{picture}
	\end{center}
	\caption[tower of states]{Physical states in the two sectors.}
	\label{tower}
\end{figure}
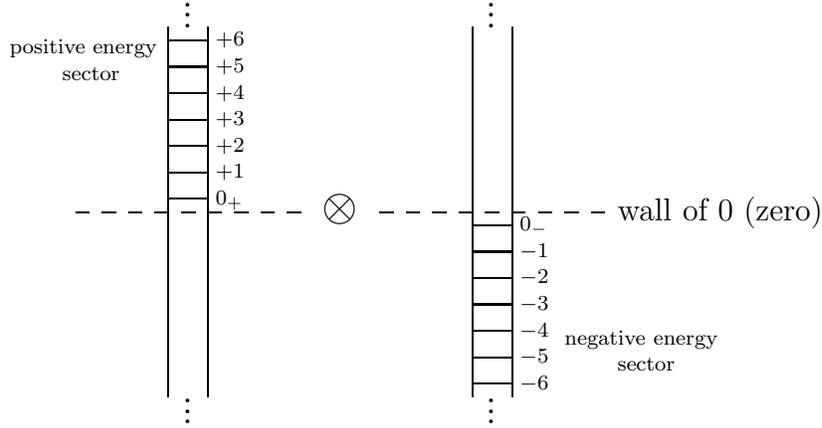

To avoid the misconceptions that the positive and negative energy sectors may simultaneously coexist and that there is no distinction between them, we depict them in Fig.~\ref{tower}.

Some explanation is in order. Figure~\ref{tower} displays towers that consist of states in the positive and negative energy sectors. In the positive energy sector, transitions upward and downward occur through the actions of $a_+^{\dagger}(\vec{k})$ and $a_+(\vec{k})$, respectively, while in the negative energy sector, $a_-^{\dagger}(\vec{k})$ and $a_-(\vec{k})$ produce transitions upward and downward, respectively. There exist positive energy and negative energy excited states in each sector, as is understood from the method of the construction of the double harmonic oscillator. However, in the positive energy sector depicted in Fig.~\ref{tower}, only the positive energy excited states have physical meaning. Similarly, in the negative energy sector, only the negative energy excited states have physical meaning. The reason is that there exists a wall, the classical number ``0 (zero)", which cannot be penetrated with the usual polynomial interaction of field theories. Consequently, we can conclude that the total vacuum of the complex scalar boson system is given by the direct product (70) of the empty positive energy sector vacuum and the modified negative energy sector vacuum, so that $a_-^{\dagger}(\vec{k})||0_-\rangle =0$, i.e. to obey (72). The negative energy sector vacuum is a state in which an infinite number of negative energy bosons, in a sense, exist, as described by Eq.(79), and furthermore, by acting with $a_-(\vec{k})$, we can annihilate an infinite number of negative energy particles. Thus, there exist all numbers (including infinity) of excited states. In fact, in the boson theory, the relation (79), which relates the empty vacuum $||0_+\rangle$ and the sea vacuum $||0_-\rangle$, exponentiated operators is preferable and, indeed, needed.

Next, we study the possible values of the energy of the boson system for which the vacuum is represented by a sea, by constructing the energy-momentum operator. This operator is derived using Noether's theorem: 

\begin{align}
	P^{\mu} & =\int d^3\vec{x}\left\{ \frac{\partial \mathcal{L}}
	{\partial (\partial_{\mu}A(x))}\partial_0A(x)
	+\frac{\partial \mathcal{L}}{\partial (\partial_{\mu}A^{\dagger}
	(x))}\partial_0A^{\dagger}(x)-\eta^{\mu 0}\mathcal{L}\right\} 
	\nonumber \\
	& =\int d^3\vec{k}\> \frac{1}{2}k^{\mu}\left\{ a_+(\vec{k})
	a_+^{\dagger}(\vec{k})+a_+^{\dagger}(\vec{k})a_+(\vec{k})+a_-(\vec{k})
	a_-^{\dagger}(\vec{k})+a_-^{\dagger}(\vec{k})a_-(\vec{k})\right\}.
\end{align}

\noindent With the commutator (69), if we define 

\begin{align}
	P^{\mu}=\int d^3\vec{k}\> k^{\mu}\left\{ a_+^{\dagger}(\vec{k})
	a_+(\vec{k})+a_-(\vec{k})a_-^{\dagger}(\vec{k})
	+\int \frac{d^3\vec{x}}{(2\pi)^3}\> 1\right\},
\end{align}

\noindent we obtain a physically consistent particle picture. For instance, the energy of the vacuum $||0\rangle_{\text{boson}}=||0_+\rangle \otimes ||0_-\rangle$ is computed as 

\begin{align}
	P^{\mu}||0\rangle_{\text{boson}} & =\int d^3\vec{k}\> k^{\mu}\left\{ 
	a_+^{\dagger}(\vec{k})a_+(\vec{k})+a_-(\vec{k})a_-^{\dagger}(\vec{k})
	+\int \frac{d^3\vec{x}}{(2\pi)^3}\> 1\right\} 
	||0_+\rangle \otimes ||0_-\rangle \nonumber \\
	& =\int \frac{d^3\vec{k}d^3\vec{x}}{(2\pi)^3}\> k^{\mu}
	||0\rangle_{\text{boson}}.
\end{align}

\noindent It is worthwhile to note that the boson vacuum energy (82) is exactly canceled by the fermion one if we consider a supersymmetric hypermultiplet. 

The energies of the first excited states are calculated similarly: 

\begin{align}
	P^{\mu}a_+^{\dagger}(\vec{p})||0\rangle_{\text{boson}} & 
	=\int d^3\vec{k}\> k^{\mu}
	\left\{ a_+^{\dagger}(\vec{k})a_+(\vec{k})+a_-(\vec{k})a_-^{\dagger}
	(\vec{k})+\int \frac{d^3\vec{x}}{(2\pi)^3}\> 1\right\} 
	a_+^{\dagger}(\vec{p})||0\rangle_{\text{boson}} \nonumber \\
	& =\left\{p^{\mu}+\int \frac{d^3\vec{k}d^3\vec{x}}{(2\pi)^3}\> k^{\mu}
	\right\} a_+^{\dagger}(\vec{p})||0\rangle_{\text{boson}} , \\
	P^{\mu}a_-(\vec{p})||0\rangle_{\text{boson}} & =\int d^3\vec{k}\> 
	k^{\mu}\left\{ a_+^{\dagger}(\vec{k})a_+(\vec{k})+a_-(\vec{k})
	a_-^{\dagger}(\vec{k})+\int \frac{d^3\vec{x}}{(2\pi)^3}\> 1\right\} 
	a_-(\vec{p})||0\rangle_{\text{boson}} \nonumber \\
	& =\left\{p^{\mu}+\int \frac{d^3\vec{k}d^3\vec{x}}{(2\pi)^3}\> k^{\mu}
	\right\} a_-(\vec{p})||0\rangle_{\text{boson}}.
\end{align}

\noindent From these results, it is seen that the energy-momentum of the states excited under the action of $a_+^{\dagger}(\vec{p})$ and $a_-(\vec{p})$ are increased by an amount $p^{\mu}$. This implies the following. In the positive energy sector, the interpretation is the usual one, in which one particle of energy-momentum $p^{\mu}$ is created on top of the empty vacuum. In the negative energy sector, because in the vacuum state an infinite number of negative energy particle states are filled, when $a_-(\vec{p})$ acts on this state, one negative energy particle is annihilated, creating a hole that is interpreted as an anti-particle, because this hole is considered to have energy-momentum $p^{\mu}$ relative to the surrounding particles with negative energy-momentum. All other higher states can be interpreted in the same manner. Therefore, defining the energy-momentum operator $P^{\mu}$ by Eq.(81), we obtain a physical particle picture that is consistent with that of the second quantization, in which all the excited states have positive energies. 

To end the present section, a comment about the inner product of the states in the second quantized theory is in order. If we write the $n$-th excited state as $||n\rangle \simeq \Phi_n[A]$, the inner product is defined by 

\begin{align}
	\langle n||m\rangle =\int \mathfrak{D}A\> \Phi_n^{\dagger}[A]\Phi_m[A],
\end{align}

\noindent where on the right-hand side there appears a functional integration over the scalar field $A(\vec{k})$ and $\Phi_n^{\dagger}[A]$ means $\left[ \Phi_n[A^{\ast}]\right]^{\ast}$ as in the single particle system in Section 3. Recalling the definition of a convergent inner product in the first quantization, (47), it might be thought that the integration variable $A$ should be replaced by $iB$ for $n,m=0_-,-1,-2,\cdots$, in order to make the integration convergent.

\section{Conclusion and future perspective}

\vspace{0.5cm}

In this paper, we have proposed the idea that the boson vacuum forms a sea, like the Dirac sea for fermions, in which all the negative energy states are filled. This was done by introducing a double harmonic oscillator, which stems from an extension of the concept of the wave function. Furthermore, analogous to the Dirac sea where due to the exclusion principle each negative energy state is filled with one fermion, in the boson case we also discussed a modification of the vacuum state so that one could imagine two types of different vacuum fillings for all the momenta. In the picture of the double harmonic oscillator, we imagine the analogy of filling the Dirac sea for fermions to be to replace for each nagative single boson orbit the state $||0_+\rangle$ by $||0_-\rangle$, an operation that in~\cite{nn,nn2} were described as taking a boson away from the beginning vacuum. It must, however, be immediately admitted that this beginning vacuum $||0_+\rangle$ does not lead to an acceptable quantum field theory in the usual way, because of having negative norm square states constructed from it. The story of thinking of such a beginning is therefore only of interest for developing an intuitive and pedagogical picture. But accepting this formal extension (of the Hilbert space to an indefinite one), the usual interpretation of an anti-particle, as a hole in the negative energy sea, turns out to be applicable not only for the case of fermions but also for that of bosons. Thus, we have proposed a way of resolving the long-standing problem in field theory that the bosons cannot be treated analogously to the old Dirac sea treatment of the fermions. Our presentation relies on the introduction of the double harmonic oscillator, but that is really just to make it concrete. What is really needed is that we formally extrapolate to have negative numbers of particles in the single particle states, precisely what is described by our ``double harmonic oscillator", which were extended to have negative numbers of excitation quanta. Supersymmetry also plays a substantial role in the sense that it provides us with a guideline for how to develop the method. In fact, our method is physically very natural when we consider supersymmetry, which, in some sense, treats bosons and fermions on an equal footing. 

Our picture of analogy between fermion and boson sea description is summarized by Table 2.

\begin{center}
	\begin{tabular}{|c|c|c|} \hline
	& \multicolumn{2}{|c|}{\textbf{\large Fermions}} \\
	& \small{positive single particle energy} & 
	\small{negative single particle energy} \\
	& $E>0$ & $E<0$ \\ \hline
	& & \\
	empty $||\tilde{0}_+\rangle$ & true & not realized in nature \\
	& & \\ \hline
	& & \\
	filled $||\tilde{0}_-\rangle$ & not realized in nature & true \\
	& & \\ \hline
	\multicolumn{3}{c}{} \\ \hline
	& \multicolumn{2}{|c|}{\textbf{\large Bosons}} \\
	& \small{positive single particle energy} & 
	\small{negative single particle energy} \\
	& $E>0$ & $E<0$ \\ \hline
	``sector with bottom" & & \\
	analogous to ``empty" & true & not realized in nature \\
	$||0_+\rangle$ & & \\ \hline
	``sector with top" & & \\
	analogous to ``filled" & not realized in nature & true \\
	$||0_-\rangle$ & & \\ \hline
	\multicolumn{3}{c}{} \\
	\multicolumn{3}{c}{Table 2: Analogy between fermion and boson sea 
	description} \\
	\end{tabular}
\end{center}

Our method for constructing the boson sea is expected to have a wide range of applications in quantum physics. First of all, string theories and string field theories have been successfully quantized using the light-cone quantization method~\cite{kaku}. In particular, for string field theories there are no satisfactory theories at the present time. Our present task is to clarify why only the light-cone quantization can be carried out. In other words, we would like to determine the essential points that make covariant quantization so difficult. In attacking these problems, our method may be useful, because we believe that perhaps the reason why only the light-cone quantization is successful is that it removes all the negative energy states. Therefore, to carry out the covariant quantization, we should treat the negative energy states in a sophisticated manner. Our method may be useful for this purpose. This is one of the main motivations to develop our method of treating the old problem of quantizing bosons by appropriately including negative energy states, in analogy to the Dirac's method for fermions.

\vspace{0.5cm}

\noindent \underline{ Acknowledgement }

\vspace{0.25cm}

Y.H. is supported by The 21st century COE Program ``Towards a New Basic Science; Depth and Synthesis", of the Department of Physics, Graduate School of Science, Osaka University. This work is supported by Grants-in-Aid for Scientific Research on Priority Areas, Number of Areas 763, ``Dynamics of strings and Fields", from the Ministry of Education of Culture, Sports, Science and Technology, Japan.

\vspace{0.5cm}



\begin{thebibliography}{nn}
\addtolength{\itemsep}{-5pt}
\bibitem{weinberg} S. Weinberg, ``The Quantum Theory of Fields, Volume I: Foundations", Cambridge University Press
\bibitem{dirac} P. A. M. Dirac, Proc. Roy. Soc. A126, 360 (1930); P. A. M. Dirac, Proc. Camb. Phil. Soc. 26, 361 (1930)
\bibitem{nn} H. B. Nielsen and M. Ninomiya, Phys. Lett. B130, 389 (1983)
\bibitem{nn2} H. B. Nielsen and M. Ninomiya, hep-th/9808108
\bibitem{sohnius} M. F. Sohnius, Physics Reports (Review Section of Physics Letters) 128, Nos. 2 \& 3 (1985) 39-204
\bibitem{west} P. West, ``Introduction to Supersymmetry and Supergravity", World Scientific
\bibitem{kaku} M. Kaku and K. Kikkawa, Phys. Rev. D10, 1110 (1974)
\end{thebibliography}
\end{document}